\RequirePackage{fix-cm}
\documentclass[smallextended]{svjour3}       
\smartqed  
\usepackage{graphicx}
\begin{document}

\title{Simulation of Light Antinucleus--Nucleus Interactions\thanks{On behalf of the Geant4 Hadronics Working Group}
}


\author{A. Galoyan         \and
        V. Uzhinsky 
}


\institute{A. Galoyan \at
              LHEP, JINR, Dubna, Russia \\
              Tel.: +7496-21-64049\\
              \email{fauthor@example.com}           
           \and
           V. Uzhinsky \at
              LIT, JINR, Dubna, Russia and CERN, Geneva, Switzerland
}

\date{Received: date / Accepted: date}

\maketitle

\begin{abstract}
Creations of light anti-nuclei (anti-deuterium, anti-tritium, anti-He3 and anti-He4) are observed
by collaborations at the LHC and RHIC accelerators. Some cosmic ray experiments are aimed to find the
anti-nuclei in cosmic rays. To support the experimental studies of the anti-nuclei
a Monte Carlo simulation of anti-nuclei interactions with matter is implemented in the {\sc Geant4}
toolkit. The implementation combines practically all known theoretical approaches to the problem of
antinucleon-nucleon interactions.

\keywords{anti-nucleus \and anti-proton \and cross sections \and 
annihilation  \and quark-gluon string model }
\PACS{25.43.+t \and 13.75.Cs \and 21.60.Ka \and 25.45.-z \and 25.45.De}
\end{abstract}

\section{Introduction}
\label{intro}
One of the most exciting puzzles in cosmology is connected with the question of the existence
of anti-matter in the Universe. A number of dedicated cosmic ray experiments aim to search for anti-nuclei
\cite{PAMELA,BESS,AMS,CAPRICE}.  Also, anti-nuclei have been observed in nucleus-nucleus and
proton-proton collisions  by experiments at the RHIC \cite{STAR,PHENIX,PHOBOS} and LHC
accelerators \cite{ALICE}. The STAR collaboration at RHIC reported in March 2011 that
the anti-He4 nuclei were identified in high energy nucleus-nucleus collisions
(see \cite{StarAHe4}). The ALICE collaboration at LHC confirmed the STAR
results in May 2011.

An experimental study of anti-nuclei requires a knowledge of anti-nucleus
interaction cross sections with matter. The cross sections are needed to estimate various
experimental corrections, especially those due to particle losses which reduce the detected
rate. In practice, various phenomenological approaches are applied in order to estimate
the antinucleus-nucleus cross sections. Thus, a first task is a creation of reliable estimations
of the cross sections. Here we use the Glauber approach.

It is obvious that an annihilation can take place at an interaction of an anti-nucleus with a nucleus.
A lot of mesons can be produced in this way. Thus, we have to simulate the meson production. We do this
in the framework of the quark-gluon string model.

Low energy mesons can have secondary interactions in nuclear residues. We take them into account using
the binary cascade model of the {\sc Geant4} toolkit \cite{GEANT4}.

\section{Antinucleus-nucleus cross sections}
\label{sec:1}
Anti-proton elastic scattering by deuterons was considered in the classic paper by V. Franco
and R.J.~Glauber \cite{Franco_Glauber}. O.D. Dalkarov and V.A. Karmanov \cite{Dalk_Karm}
showed that elastic and inelastic (with excitation of nuclear levels) anti-proton scattering
by C, Ca, and Pb nuclei are described quite well at $\bar p$ kinetic energies of 46.8 and
179.7 MeV.
The first calculations of the cross sections of anti-deuteron interactions with nuclei in
the eikonal approximation were presented by Buck et al. \cite{AdAeikonal} (see also
\cite{China}). Cross sections of antideuteron-deuteron interactions at $P_{\bar d}=$ 12.2
GeV/c were calculated using the Glauber approach in \cite{Ludmila}. They were in good
agreement with the experimental data. We use the Glauber approach
to calculate the antinucleus-nucleus cross sections.

The elastic scattering amplitude of an anti-nucleus containing $\bar{A}$ anti-baryons on
a target nucleus with mass number $A$ is given as \cite{Franco68}:
\begin{equation}
F_{\bar{A}A}(\vec q) =\frac{i}{2\pi}\ \int d^2b\ e^{i\vec q \vec b}\
\left\{ 1-\prod^{\bar{A}}_{i=1}\prod^A_{j=1}\left[ 1- \gamma(\vec b + \vec \tau_i - \vec s_j)\right]
\right\}|\Psi_{\bar{A}}|^2 |\Psi_A|^2\ \cdot
\label{eq1}
\end{equation}
$$
\left( \prod^{\bar{A}}_{i=1}d^3\ t_i\right)\
\left( \prod^A_{j=1}d^3\ r_j\right)=i\ \int^{\infty}_0 bP_{\bar{A}A}(b)\ J_0(qb)db,
$$
where the same nomenclature as in \cite{Franco68,Dalk_Karm} is used. 

The main ingredients of the approach are parameterizations of energy dependencies of total and
elastic antinucleon-nucleon scatterings. We have considered the question in our recent publication
\cite{OurPaperPL}. Using them, one can calculate various cross sections
of antiproton-nucleus and antinucleus-nucleus interactions.

\begin{figure}
  \includegraphics[width=13pc]{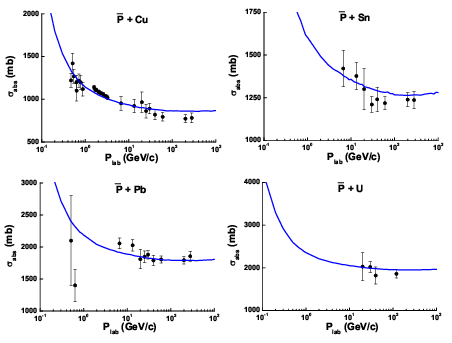}\hspace{2pc}\includegraphics[width=13pc]{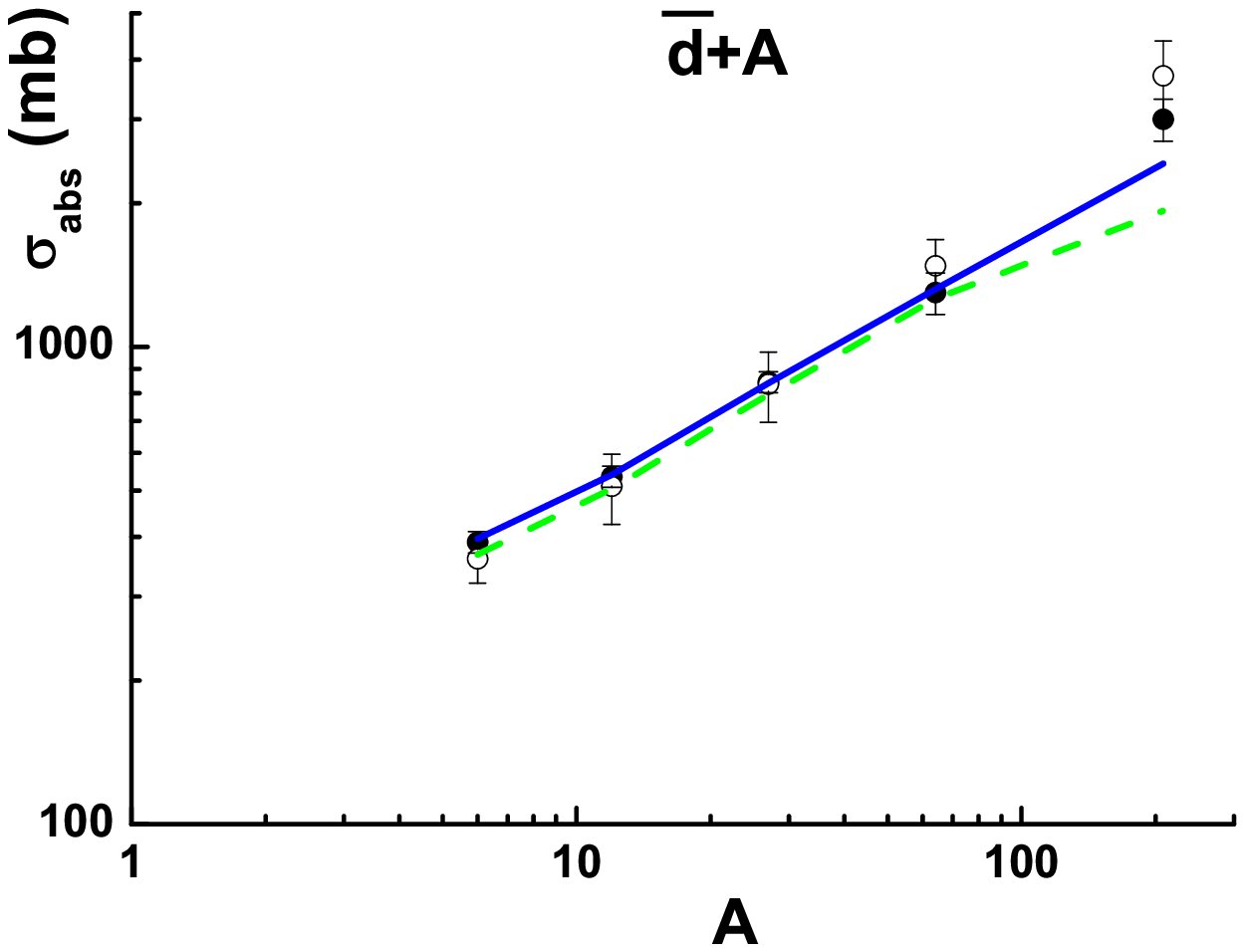}
  \caption{Absorption cross sections of anti-proton and anti-deuteron interactions with nuclei.
                      The points are experimental data \protect{\cite{Denisov}}, the lines are our calculations.}
\label{fig:1}       
\end{figure}

We show in Fig. \ref{fig:1} our calculations in a comparison with experimental data. For projectile
anti-deuterons we present data at two energies (\cite{Denisov})
-- 13.3 and 25 GeV/c (open and close circles, correspondingly) and calculation results at the energies
(solid and dashed lines). As seen, the agreement between the experimental data and the calculations is
rather good.

The profile function of the elastic antinucleus-nucleus scattering, $P_{\bar{A}A}(b)$, is very close to
the 2-dimensional Fermi-function, $P_{\bar{A}A}(b)=1/[1+exp((b-R)/c)]$. The Fourier-Bessel transform of
the function gives the elastic scattering amplitude in the momentum representation. We use analytical
expressions for an expansion of the transform  presented in (\cite{Ftransform}) to calculate differential
elastic antinucleus-nucleus cross sections.  

Calculations of the total, inelastic and elastic cross-sections of
of an anti-nucleus interactions with nuclei allow to be determined a point 
where the anti-nucleus penetrating through a matter will interact.

\section{Simulation of multiparticle production in antinucleon-nucleon interactions}
\begin{figure}[cbth]
  \includegraphics[width=28pc]{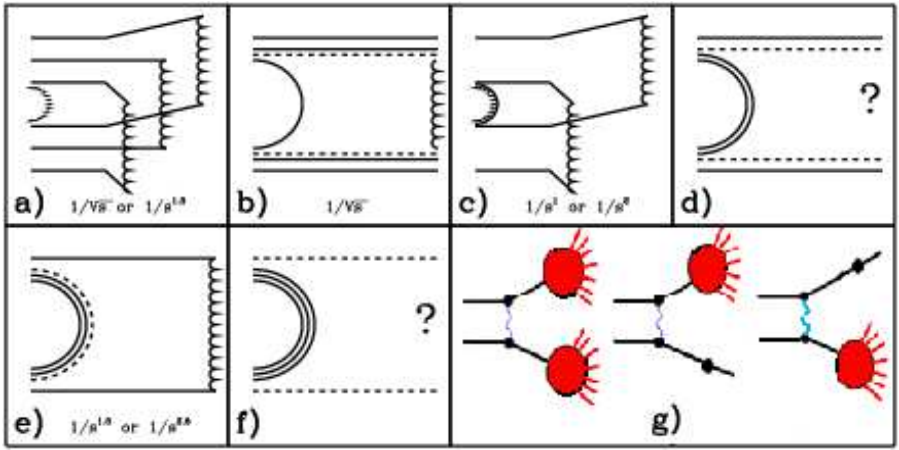}
  \caption{Quark flow diagram.}
\label{fig:2}
\end{figure}

The main channel of the antinucleon-nucleon interactions at low energies is the annihilation into 3, 4,
or 5 mesons. It is commonly assumed that the reaction is going through the re-arangement of quarks and
anti-quarks in the colliding particles, see Fig. \ref{fig:2}a. At high energies, 
the reaction will  result in a three quark-gluon string creation.

If a quark and an antiquark annihilate (Fig. \ref{fig:2}b), diquark-antidiquark string will be left.
Such a string must fragment into a final state containing a baryon and an anti-baryon. A description
of the fragmentation of such strings is one of the problems in high energy physics. We have solved
it using experimental data on the reactions $\bar pp \rightarrow \bar nn,
\bar \Lambda \Lambda, \bar nn\pi^0$, and so on.

A string junction couples quarks in a baryon. String junction's annihilation takes place 
in the process of Fig. \ref{fig:2}a which results in a three string creation. 

At annihilation of quark, anti-quark and string junctions,  
 two strings will be created (see Fig. \ref{fig:2}c). 
 If the energy is sufficiently low, two mesons will be produced.

The diagrams "d" and "e" describe the processes with one string creation. The ordinary quark-antiquark
string can fragment at least into two mesons, and it can be responsible for the reactions with two final
state mesons. A hybrid meson with hidden baryon number can be produced in process "d" at low energies.

It is needed to add the "standard" Fritiof model \cite{Fritiof} diagrams "g" at high energies. One or two
strings can be created in the corresponding processes without baryon number exchanges.

The main issue of the considered approach is the energy dependencies of the process cross sections.
We did not find an acceptable solution of the question. Thus, we parameterized the cross sections as follow:
\begin{equation}
\sigma_a=\frac{16}{\sqrt{s-4m^2}}\left[ (s-4m^2)^{-0.175}+3.125*(1-1.88/\sqrt{s})\right]\ \ \ (mb)
\end{equation}
\begin{equation}
\sigma_b=3.13+140\cdot (M_{th}-\sqrt{s})^{2.5}
\ \ \ (mb), \ \ \ \sqrt{s}\leq M_{th}=2.172 \ \ \ (GeV)
\end{equation}
$$ \sigma_b=6.8/\sqrt{s} \ \ \ (mb),
\ \ \ \ \ \ \ \ \ \ \ \ \ \ \ \ \ \ \ \ \ \ \ \ \ \ \ \ \  \sqrt{s} > M_{th}=2.172 \ \ \ (GeV)$$
\begin{equation} \sigma_e=23.3/s \ \ \ (mb) \end{equation}
\begin{equation} \sigma_c=\sigma_d=\sigma_f =0\end{equation}
\begin{equation}\sigma_g=35*(1.-2.1/\sqrt{s}) \ \ \ (mb)\end{equation}

\begin{figure}[h]
\includegraphics[width=8pc]{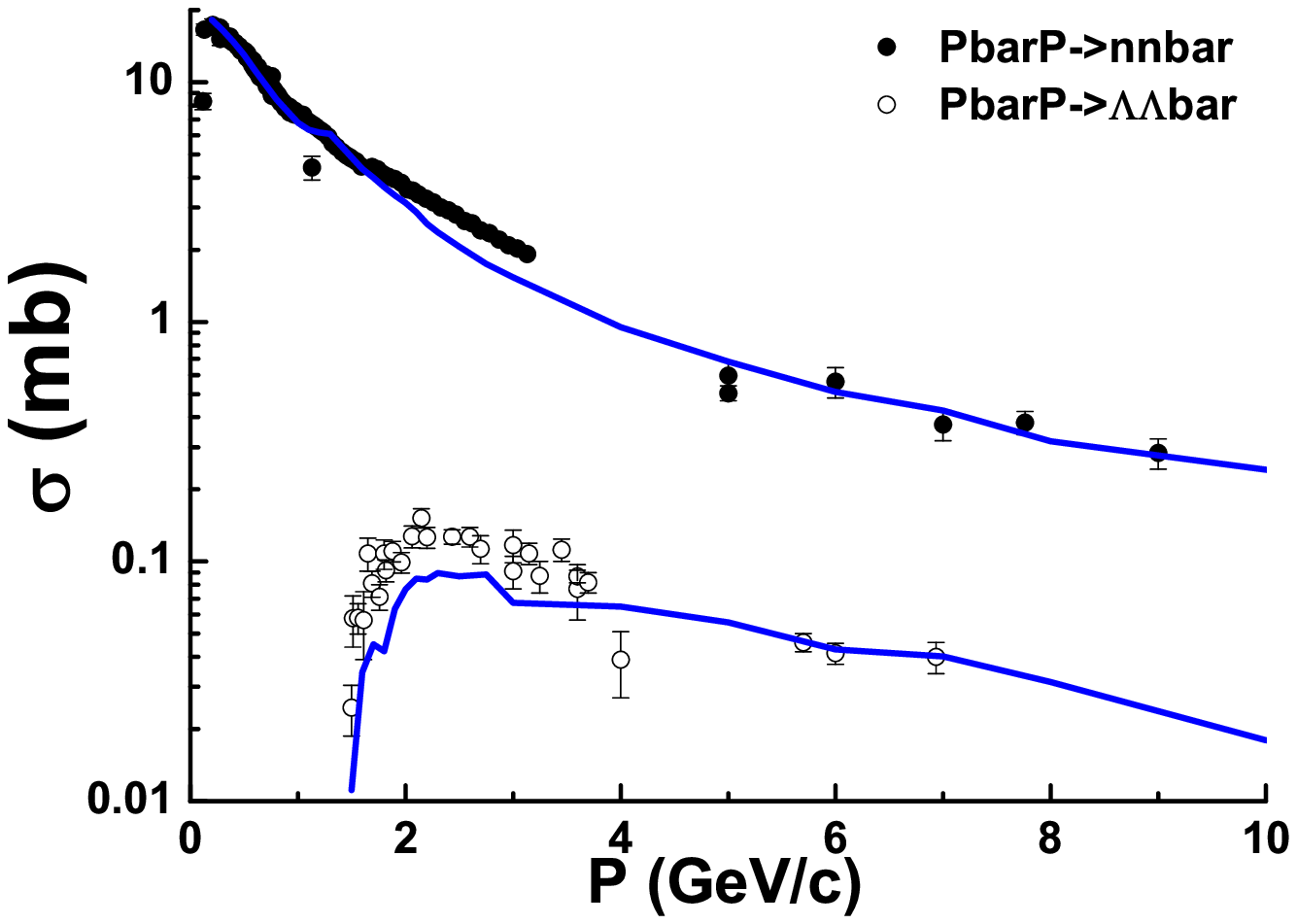}
\hspace{1pc}%
\includegraphics[width=8pc]{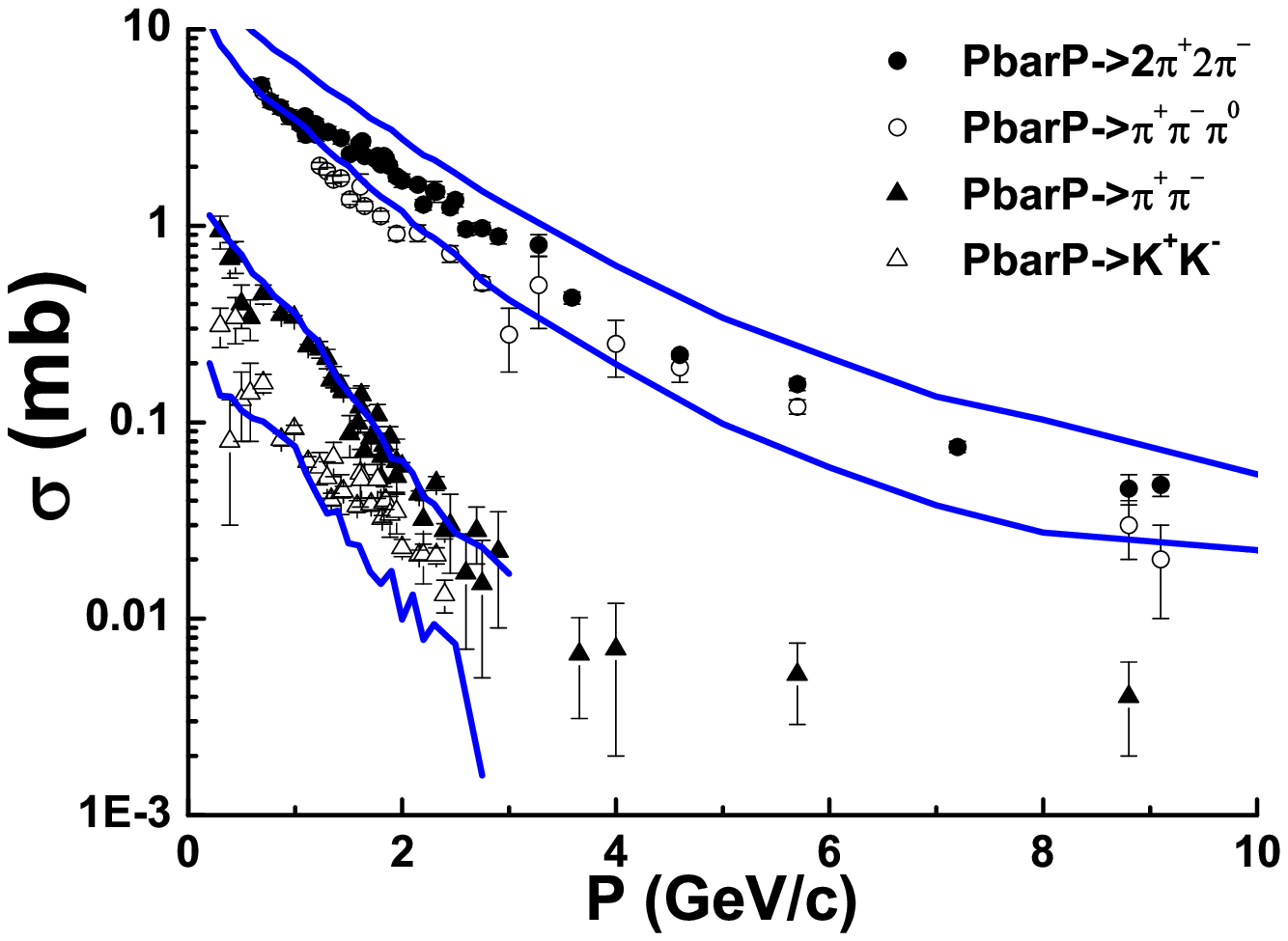}
\hspace{1pc}%
\includegraphics[width=8pc]{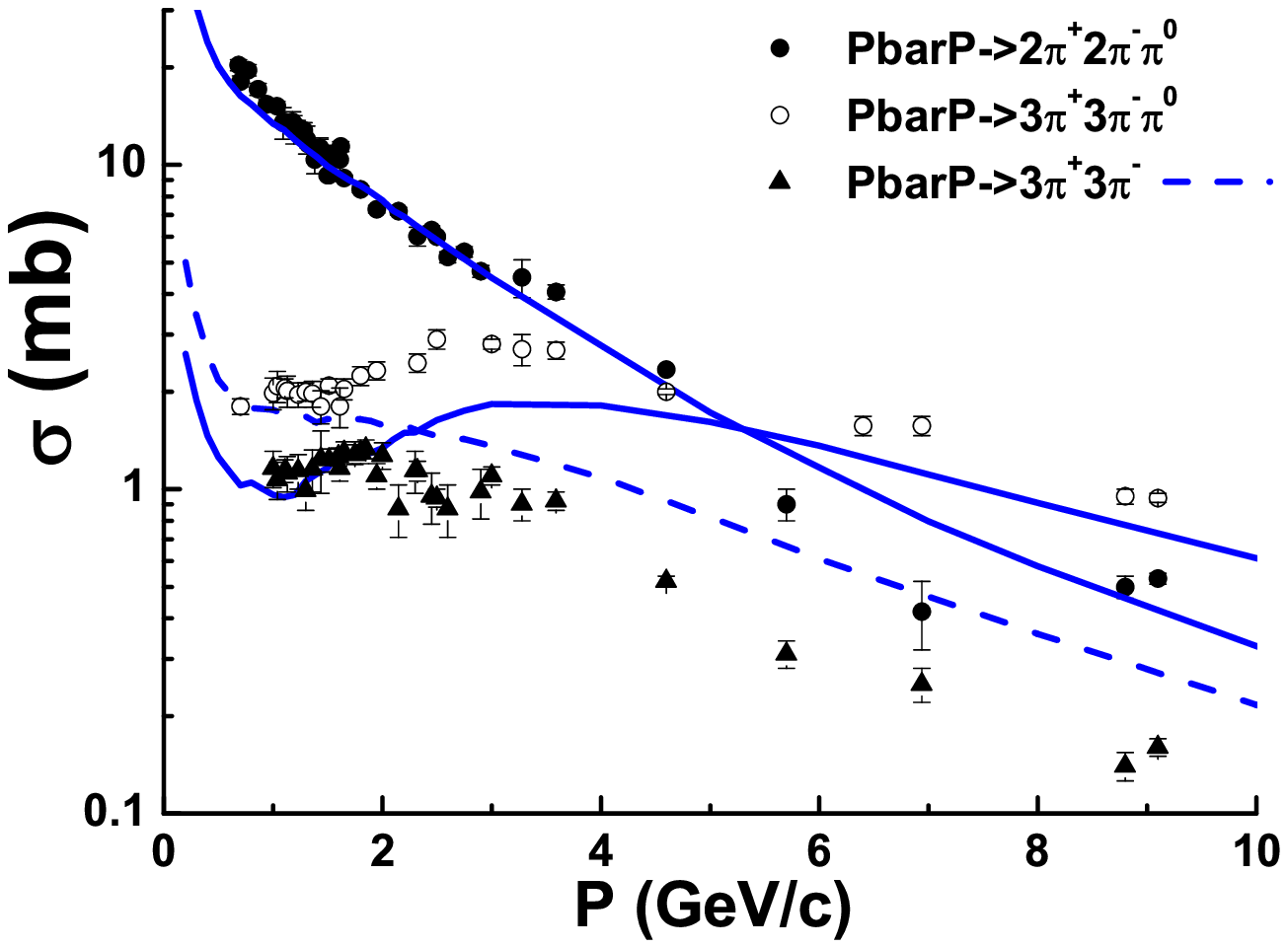}\\
\includegraphics[width=8pc]{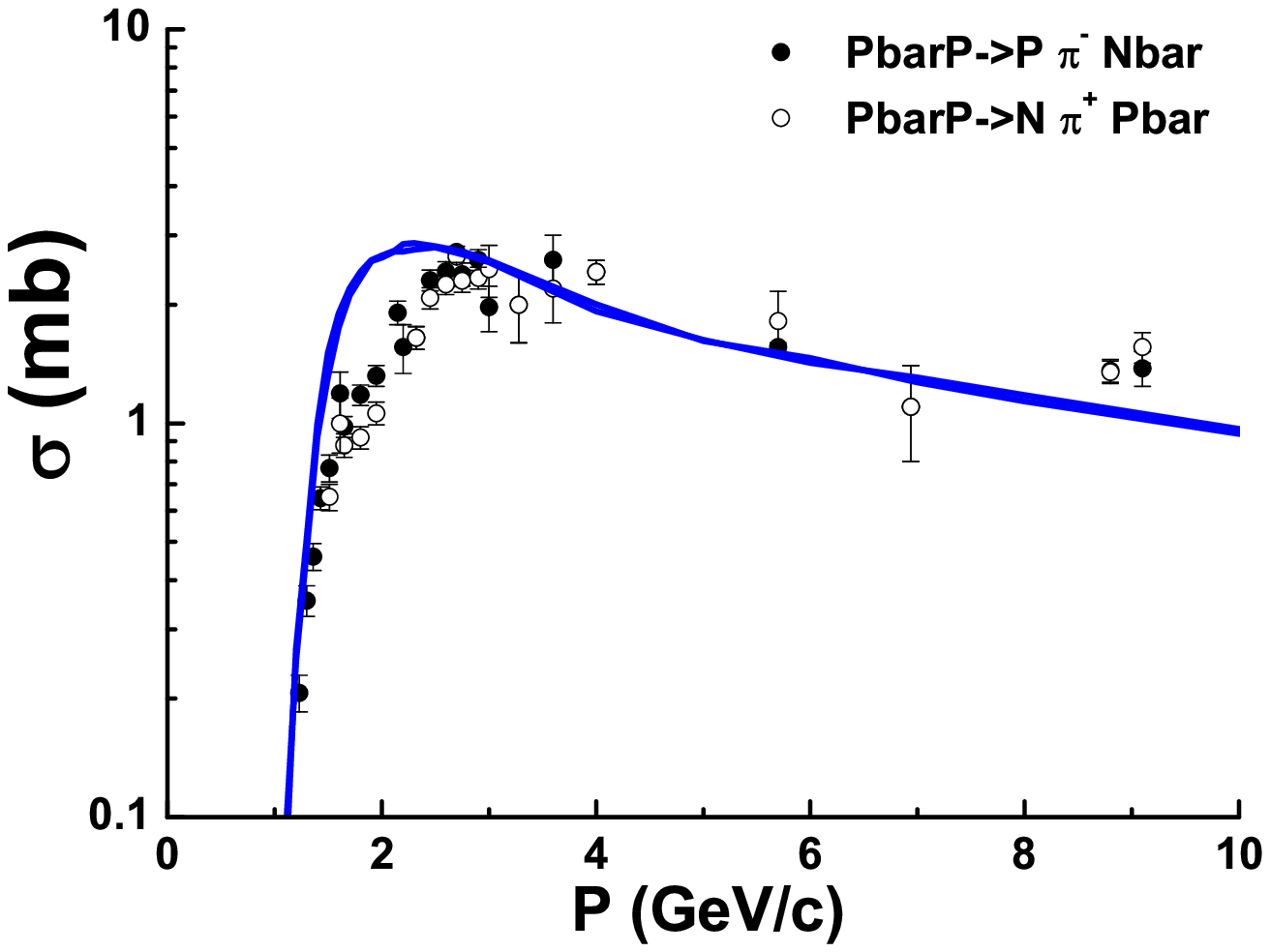}
\hspace{1pc}%
\includegraphics[width=8pc]{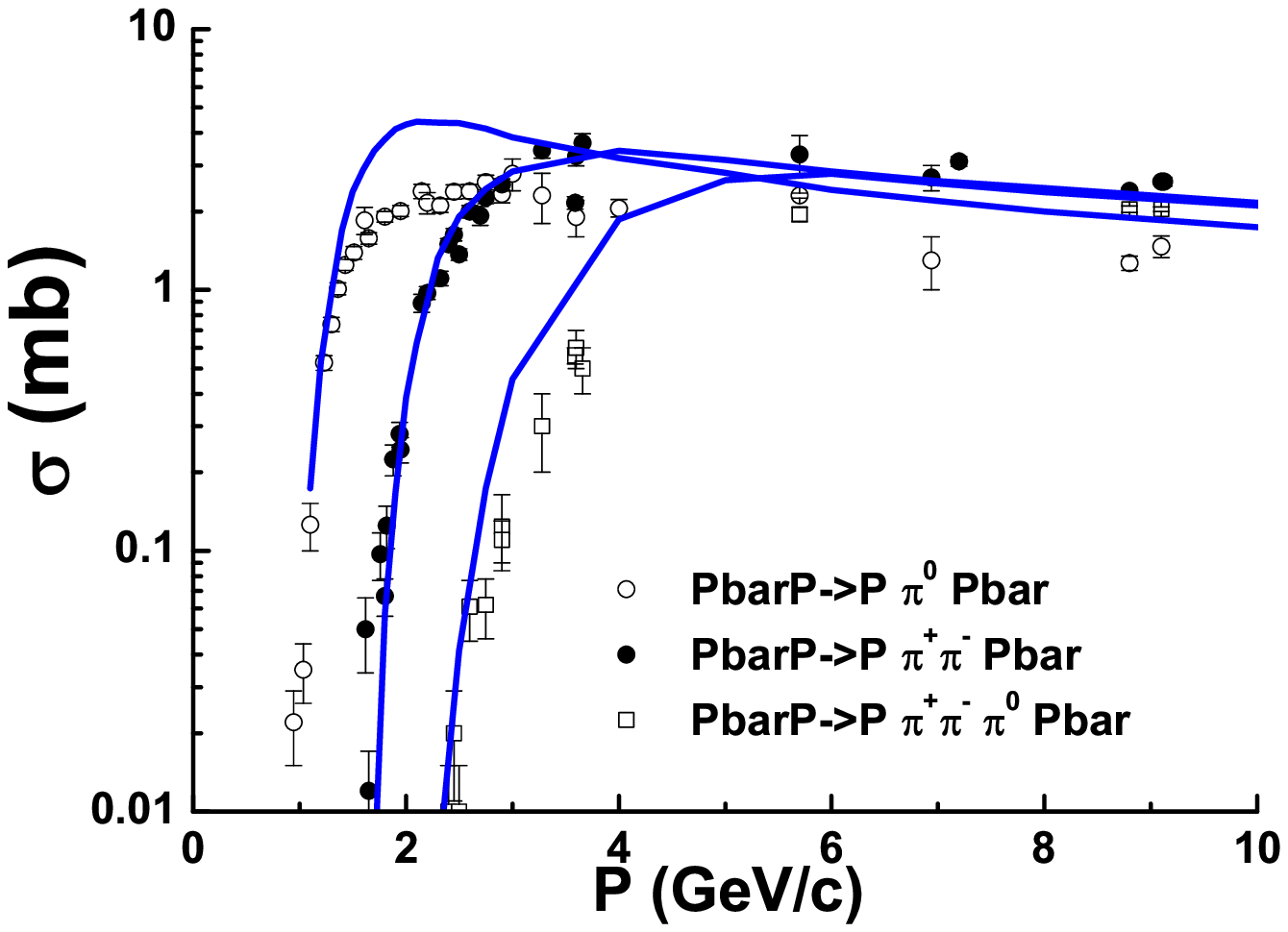}
\hspace{1pc}%
\includegraphics[width=8pc]{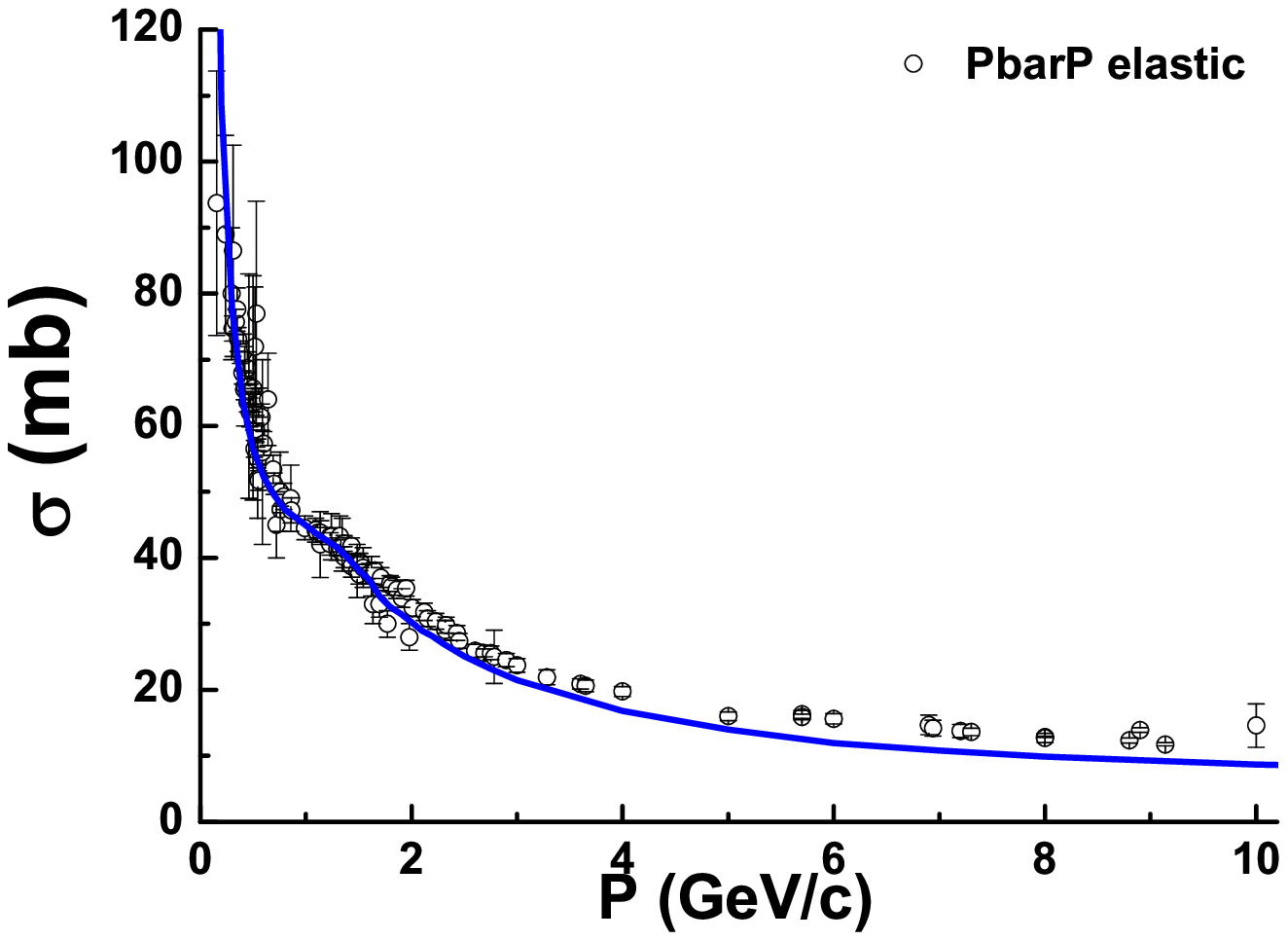}
\caption{Reaction cross sections of $\bar pp$-interactions. The points are experimental data,
             the lines are our calculations.}
\label{fig:3}
\end{figure}

The parameterizations of the cross sections and  LUND string fragmentation algorithm
are implemented in {\sc FTF} generator  of the {\sc Geant4} toolkit \cite{GEANT4}.   
Using this  generator,  we obtain results
presented in Fig. \ref{fig:3}. As seen, the experimental dependencies are reproduced, though
a more refined consideration is needed.

\section{Antiproton-nucleus interactions}
Cross sections ($\sigma_{\nu}$) of various processes with different multiplicities ($\nu$) of 
involved nuclear nucleons can be determined using the asymptotic Abramovsky-Gribov-Kancheli (AGK) cutting 
rules at high energies in application to elastic scattering amplitude: 
\begin{equation}
\sigma^{in}_{\bar{p}A}=\sum^A_{\nu=1} \sigma_{\nu},\ \ \
\sigma_{\nu}=C^{\nu}_A \int d^2b \left[ \frac{1}{A} \int g(\vec b - \vec s)
\rho_{A}(\vec s,z) d^2s\ dz\right]^{\nu} \cdot
\end{equation}
$$
\cdot \left[1- \frac{1}{A} \int g(\vec b - \vec s)\rho_{A}(\vec s,z) d^2s\ dz\right]^{A-\nu}, \ \ \
g(\vec b)=\gamma(\vec b)+\gamma^*(\vec b)-\gamma(\vec b) \cdot \gamma^*(\vec b).
$$
\begin{figure}[h]
\includegraphics[width=13pc]{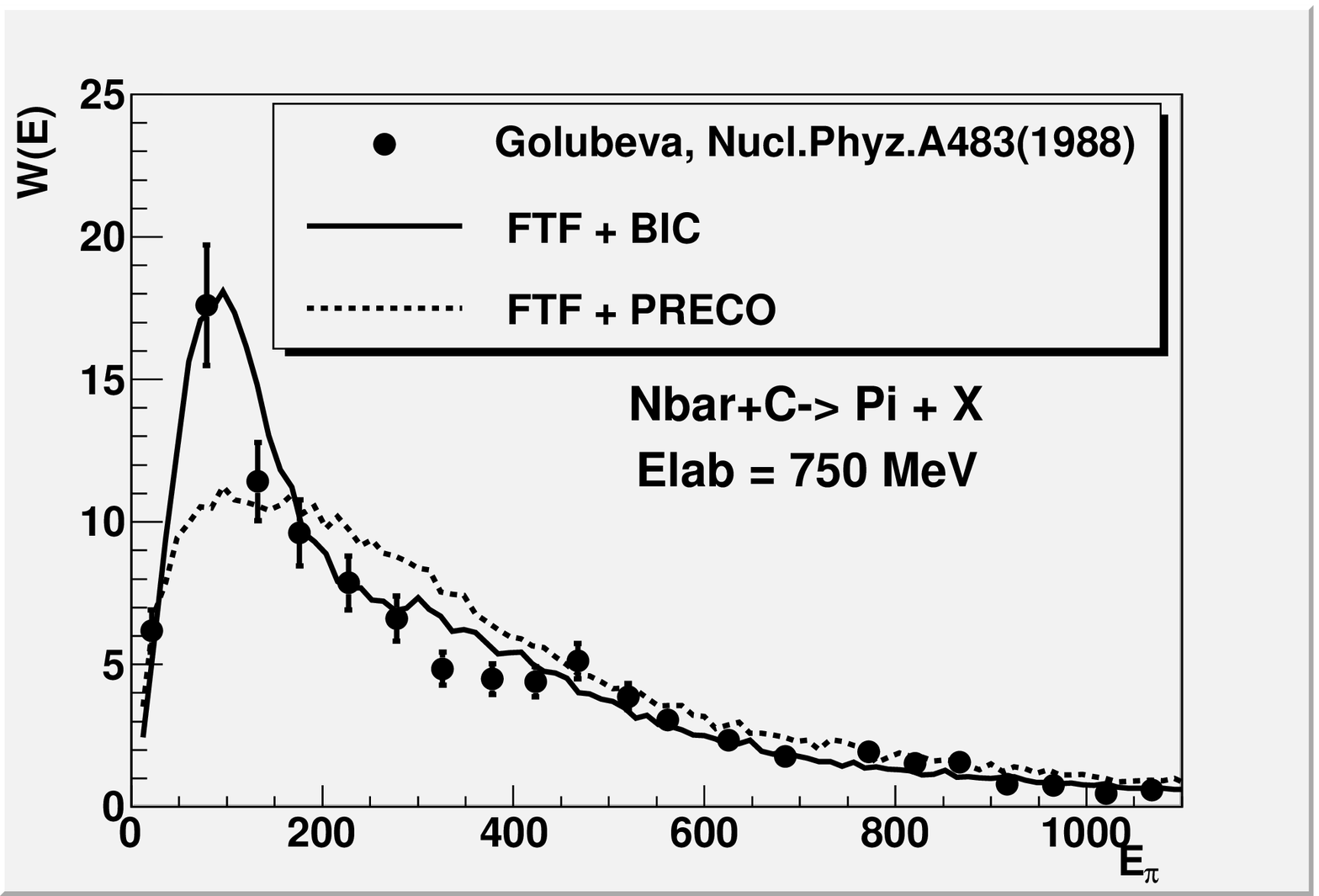}
\hspace{1pc}%
\includegraphics[width=13pc]{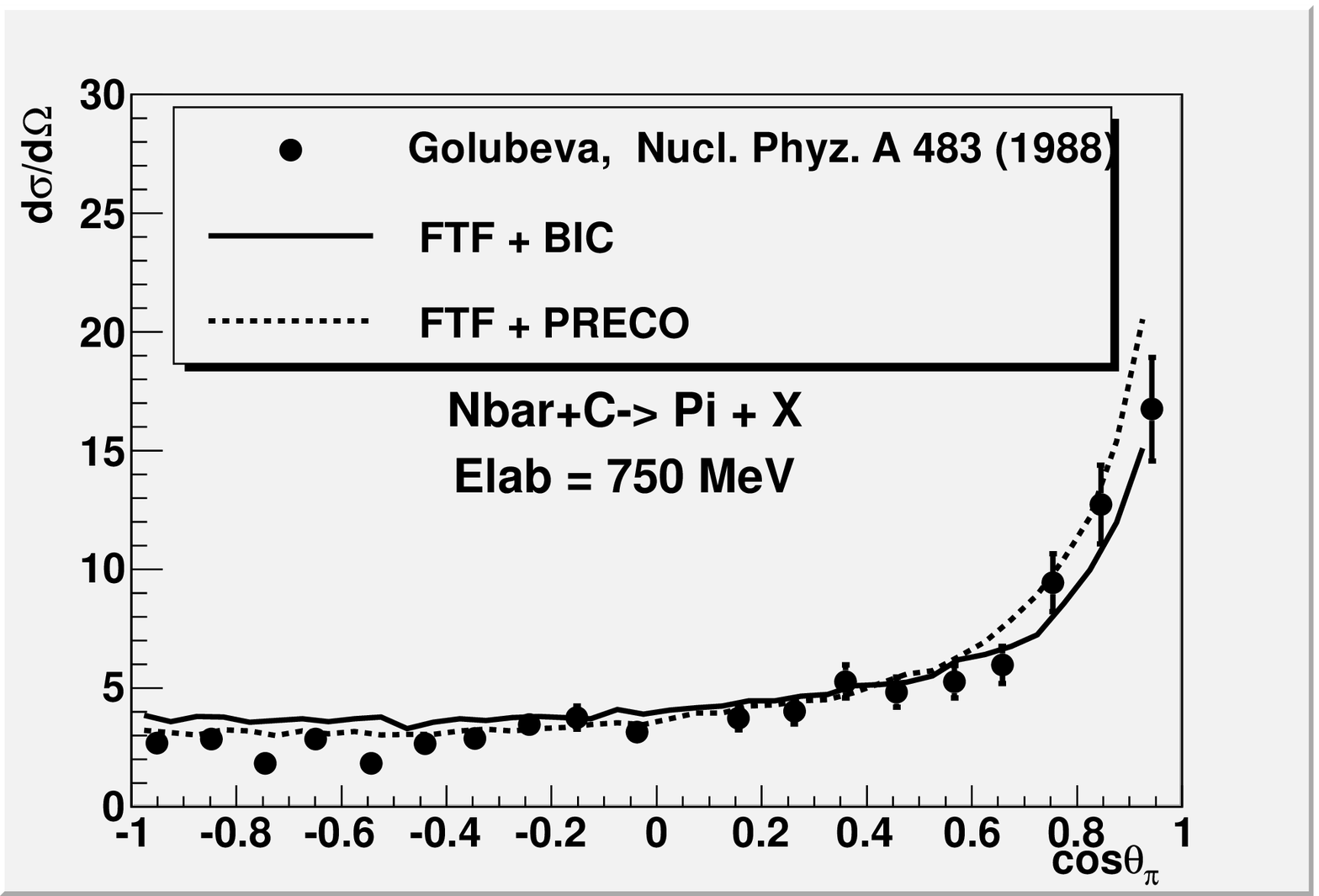}\\
\includegraphics[width=13pc]{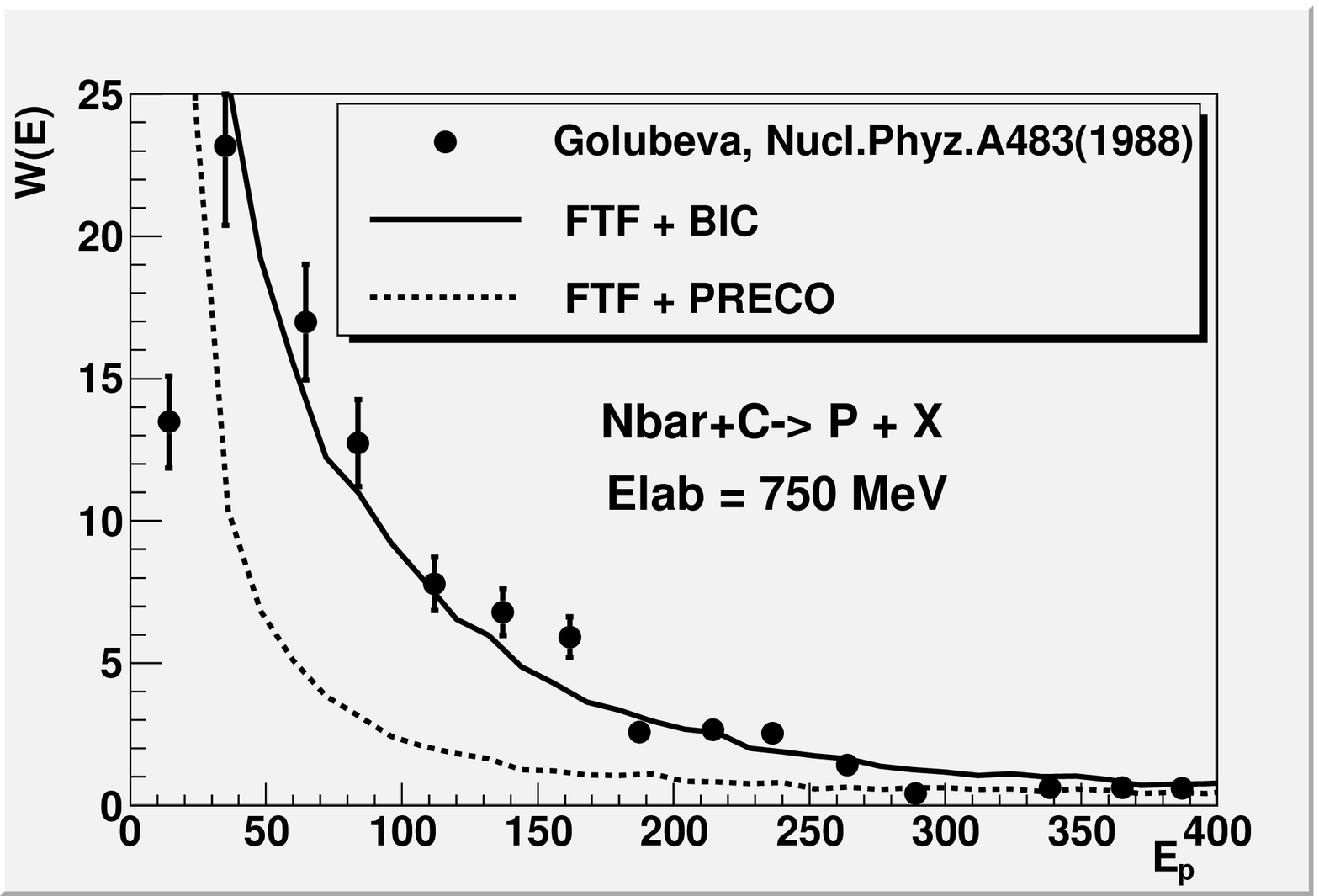}
\hspace{1pc}%
\includegraphics[width=13pc]{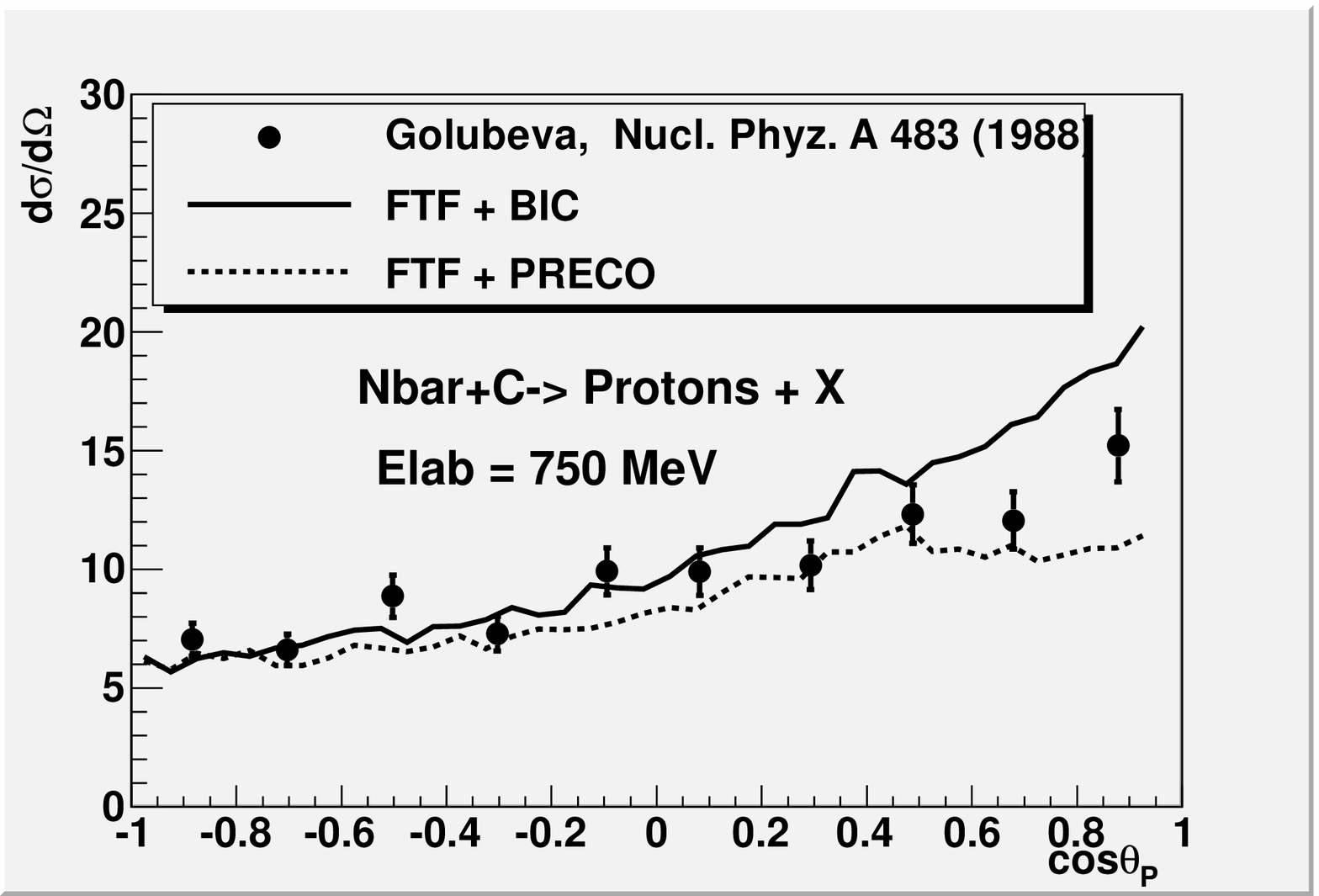}\\
\includegraphics[width=13pc]{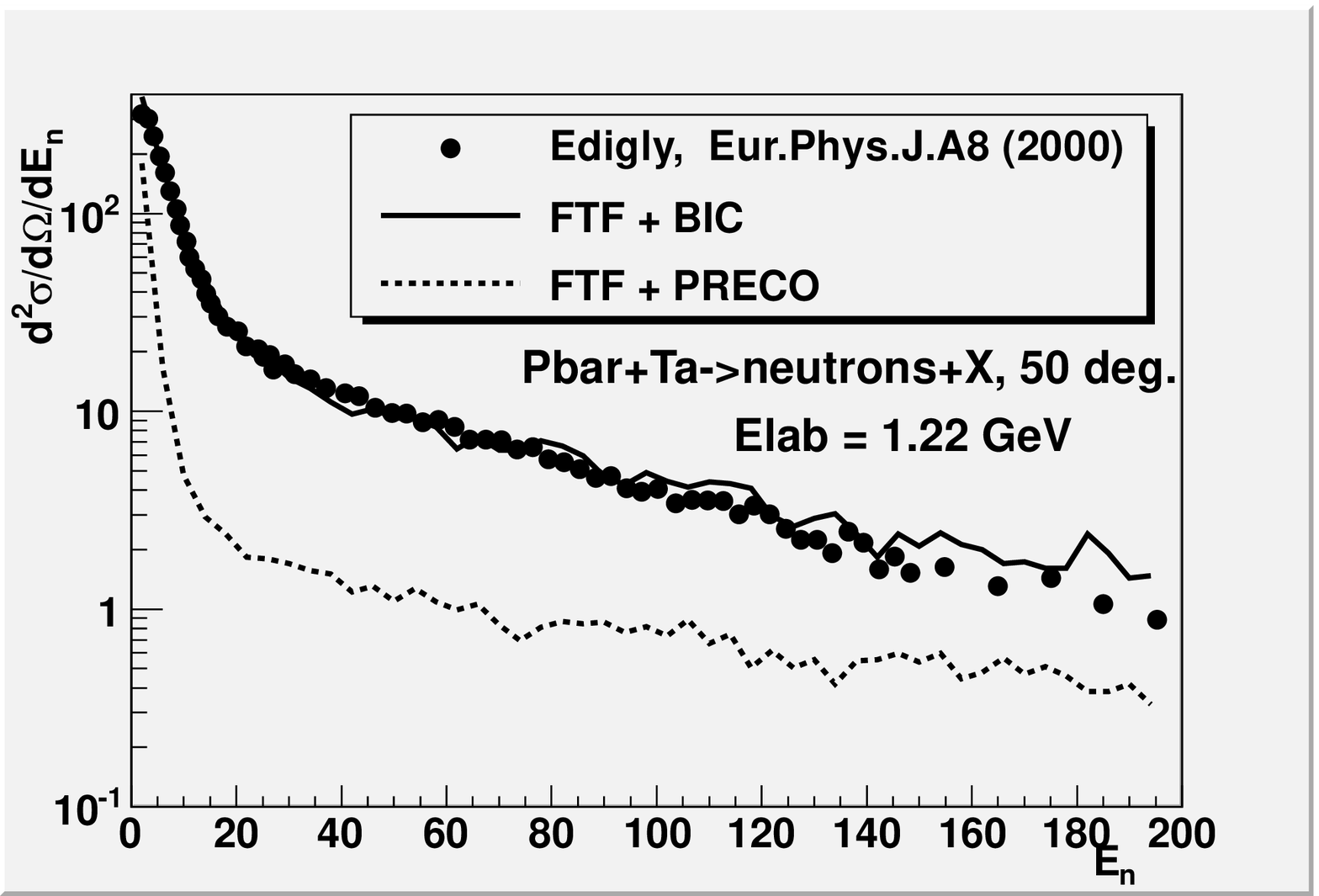}
\hspace{1pc}%
\includegraphics[width=13pc]{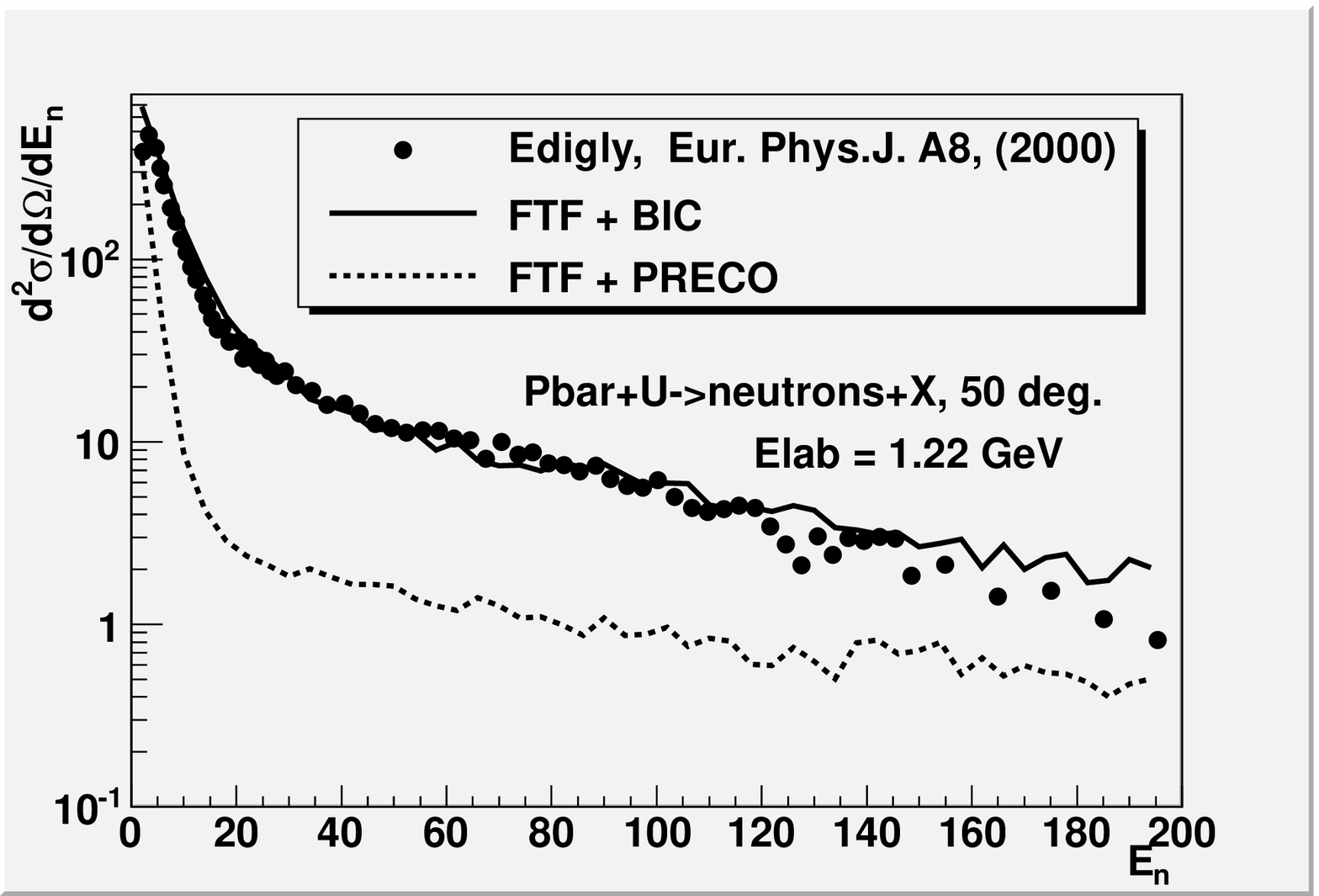}
\caption{Inclusive  $\pi^+$-meson, proton and neutron   
distributions in $\bar pA$- and $\bar{n}A$-interactions.
The points are experimental data \protect{\cite{Golub,Gold}}, the lines are 
our 
calculations.}
\label{fig:4}
\end{figure}

At low energies, we have introduced finite energy corrections to the AGK-cutting rules
and obtained 
\begin{equation}
\sigma^{in}_{\bar{p}A}=\sum^{\nu_{max}}_{\nu=1} \sigma_{\nu}',\ \ \
\nu_{max}=[p_{lab}/p_0]+1, \ \ \ p_0\simeq 2 \ (GeV/c).
\end{equation}
\begin{equation}
\sigma_{\nu}'=C^{\nu}_{\nu_{max}} \int d^2b
\left\{1-\left[1- \frac{1}{A} \int g(\vec b - \vec s)
\rho_{A}(\vec s,z) d^2s\ dz\right]^{A/\nu_{max}}\right\}^{\nu} \cdot
\end{equation}
$$
\cdot \left[1- \frac{1}{A} \int g(\vec b - \vec s)\rho_{A}(\vec s,z) d^2s\ dz\right]^{(\nu_{max}-\nu)A/\nu_{max}}.
$$

Thus at a low energy, only one inelastic interaction of a projectile in a nucleus can happen, but there 
can be other interactions caused by secondary particles. The secondary particle interactions are mainly
responsible for slow neutron and proton productions (see Fig. \ref{fig:4}). We have checked this using
two variants of the model -- FTF with the binary cascade model (BIC) and FTF with the precompount de-excitation model (PRECO)
\cite{BIC}. A secondary particle cascading and a nuclear residual de-excitation are considered in the binary cascade model,  
the second model does not take into account the cascading. 
 Both models BIC and PRECO are
presented in the {\sc Geant4} toolkit. As seen from a comparison of 
experimental data \cite{Golub,Gold} to the model calculations in Fig. 
\ref{fig:4}, 
the finite energy corrections and the secondary
interactions are very important for understanding experimental regularities.

\section{Antinucleus-nucleus interactions}
To simulate the distribution (8), we use the following algorithm: starting with the expression (7)
we ascribe a projectile a power $P=\nu_{max}$.
A probability of an interaction with the first nucleon is equal  $P/\nu_{max}$. The power
decreases after the interaction on unit. The probability of an interaction with the second nucleon is equal
to $P/\nu_{max}$, where $P=\nu_{max}-1$. If the second interaction has happened, the power is decreased
one more. In other case, it is left on the same level. This is applied for each possible interaction.
After an annihilation $P$ is set to zero. One can check that this leads to Exp. (9).

\begin{figure}[cbth]
  \includegraphics[width=28pc]{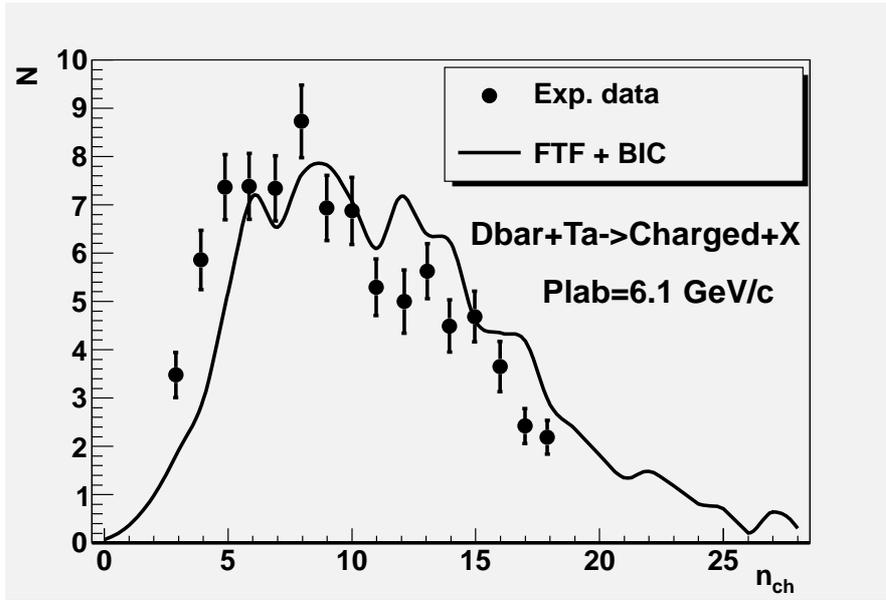}
  \caption{Charged particle multiplicity distribution in $\bar{d}+Ta$ interactions at
           $P_{\bar{d}}=6.1$ GeV/c.}
\label{fig:5}
\end{figure}

The same algorithm is applied in the case of antinucleus-nucleus interactions using the Glauber
inelastic cross sections analogous to Eq. (7). A cascading of secondary particles in the light
projectile anti-nuclei is neglected. A result of calculations is presented in Fig. \ref{fig:5},
where the points are experimental data \cite{Andreev},  the line is our calculation.

There is no other data on light anti-nucleus interactions with heavy materials except \cite{Andreev},
though there are a lot of detailed data on $\bar{d}d$- interactions, and only few data on
general properties of the interactions.

\section*{Conclusion}
Using the described approach we have developed in the {\sc Geant4} framework
a Monte Carlo model for simulation of antinucleus-nucleus interactions for the projectiles
$\bar p , \bar d , \bar t , {}^3\overline\mathrm{He} , {}^4\overline\mathrm{He}$.
The model is valid between 100 MeV/c and 1 TeV/c per antinucleon. A comparison of the
model calculations to available data shows a good agreement sufficient for most applications
in cosmic ray experiments and in large HEP experiments including those at the LHC and RHIC.

The {\sc Geant4} toolkit is now able to simulate the antinucleus-nucleus interactions
for all target nuclei since version {\it 9.4.p01}.

\begin{acknowledgements}
The authors are thankful to J. Apostolakis, G. Cosmo, G. Folger, A.~Howard, 
V.N. Ivanchenko and  D.H. Wright for stimulating
discussions and interest in the work.
\end{acknowledgements}

\bibliographystyle{spmpsci}

\end{document}